\documentclass[preprint,11pt]{aastex}
\usepackage{natbib}
\usepackage{epsfig}
\usepackage{pdfpages}
\usepackage{footnote}

\makeatletter
\renewcommand\@biblabel[1]{\hspace{-\labelsep}}
\makeatother

\begin{document}

\title{Scattering study of Pulsars below 100 MHz using LWA1}

\author{K. Bansal\altaffilmark{1}, {G. B. Taylor\altaffilmark{1}},
{Kevin Stovall\altaffilmark{2}}, \& {Jayce Dowell\altaffilmark{1}}
}

\altaffiltext{1}{Department of Physics and Astronomy, University of New Mexico, Albuquerque, NM 87131}
\altaffiltext{2}{National Radio Astronomy Observatory, Socorro, NM}

\begin{abstract}

Interstellar scattering causes pulsar profiles to grow asymmetrically, thus affecting the pulsar timing residuals, and is strongest at lower frequencies. Different Interstellar medium models predict different frequency ($\nu$) and dispersion measure (DM) dependencies for the scattering time-scale $\tau_{sc}$. For Gaussian inhomogeneity the expected scaling relation is $\tau_{sc} \propto \nu^{-4}\ DM^{2}$, while for a Kolmogorov distribution of irregularities, the expected relation is $\tau_{sc} \propto \nu^{-4.4}\ DM^{2.2}$. Previous scattering studies show a wide range of scattering index across all ranges of DM. A scattering index below 4 is believed to be either due to limitations of the underlying assumptions of the thin screen model or an anisotropic scattering mechanism. We present a study of scattering for seven nearby pulsars (DM $< 50$ pc cm$^{-3}$) observed at low frequencies ($10-88$ MHz), using the first station of the Long Wavelength Array (LWA1). We examine the scattering spectral index and DM variation over a period of about three years. The results yield insights into the small-scale structure of ISM as well as the applicability of the thin screen model for low DM pulsars. 
\end{abstract}

\section{Introduction}

The interstellar medium (ISM) consists of an ionized and turbulent plasma which causes a delay in time and variations in the phase of radio signals. The subsequent interference gives rise to a diffraction pattern and broadens the apparent size of a source. Some observable ISM effects are dispersion, scattering, angular broadening, and interstellar scintillation. The dispersion causes a delay between the pulse arrival times of the upper and lower ends of a broadband pulsar signal. The scattering of a pulse signal causes temporal broadening, making average profiles grow asymmetrically broader at lower frequencies (e.g. \citealp{lohmer01,lohmer04,lewan13,lewan15}). Pulsars are compact sources and emit short periodic pulses making them good sources for studying and understanding these propagation effects.

Assuming that the scattering occurs due to a thin screen between the observer and the source, the pulse broadening function can be expressed in terms of an exponential function $\sim exp(-t/\tau_{sc})$ with scattering parameter $\tau_{sc}$ \citep{scheuer68}. This is also known as the thin-screen model where different ISM models of electron density fluctuations predict different frequency dependencies for the scattering parameter given by $\tau_{sc} \propto \nu^{-\alpha}$, where $\alpha$ is the scattering time spectral index. This model considers an isotropic homogeneous turbulent medium. For Gaussian inhomogeneity (Cronyn 1970; Lang 1971), the scaling relation is:
\begin{equation}
\tau_{sc} \propto \nu^{-4} DM^{2},
\end{equation} 

while for a purely Kolmogorov distribution of inhomogeneities \citep{romani86}, the expected relation is:
\begin{equation}
\tau_{sc} \propto \nu^{-4.4} DM^{2.2}.
\end{equation} 
In both Equation 1 and 2, $\nu$ and DM are frequency and dispersion measure, respectively. The DM is given by $\int n_{e} dl$, where $n_{e}$ is the electron density and $dl$ is the path length along the line of sight (LOS). Measurements of DM have helped us understand the distribution of free electrons and estimate pulsar distances in our Galaxy \citep{ne2001,ne2017}.

The amount of observed scattering can be estimated by assuming a spectrum of electron density fluctuations. A simple power-law model is given by:

\begin{equation}
P_{ne}(q) = C_{ne}^{2} q^{-\beta},
\end{equation}

where $q$ is the amplitude of a three dimensional wavenumber and $C_{ne}^{2}$ is the fluctuation strength along a given LOS. The above simplification is valid when the inverse of wave-number q (1/q) is much larger than the inner scale, and much lower than the outer scale \cite{lambert99}. The value of $\beta$ ranges between 2 and 4 and is related to the scattering spectral index via:

\begin{equation}
\alpha = \frac{2\beta}{(\beta-2)} .
\end{equation}

This simplified version of the scattering strength (Equation 3) may not be valid for real cases which are difficult to predict since we do not have information about the inner/outer scales. When the diffractive scale drops below the inner scale, the dependence becomes quadratic. This change in diffractive scale with observing frequency leads to flatter spectra in comparison to the theoretical value at lower frequencies (see \citealp{rickett09,lewan15}).


In previous scattering studies, while the average value of $\alpha$ seems to agree with the theoretical models, for individual pulsars large deviations have been detected across all ranges of DM \citep{lewan15}. For large DMs ($>300$ pc cm$^{-3}$), \cite{lohmer01} report a mean value for $\alpha$ of $3.400 \pm 0.013$ obtained from frequencies between 600 MHz and 2.7 GHz. The authors explain that this could be due to the presence of multiple screens between the pulsar and the observer. \cite{lewan13,lewan15} report $\alpha$ in the range of $2.61 - 5.61$, obtained for a sample of 60 pulsars. They also see $\alpha$ values below 2.61 but discarded them due to poor data quality. \cite{geyer16} simulated anisotropicaly scattered data and fit it with the isotropic model which results in $\alpha$ values less than the theoretically predicted values as well as the effect of non-circular scattering screens leading to low $\alpha$ values ($\sim2.9$). Scattering spectra with $\alpha < 4$ have been interpreted as a limitation of assumptions underlying the thin scattering model. Plausible deviations from the thin screen assumptions include a truncated scattering screen \citep{cordes01}, the impact of an inner cut-off scale \citep{rickett09}, and anisotropic scattering mechanisms \citep{stine01}.

Moreover, ISM scattering accounts for one of the largest time-varying sources of noise in timing residuals of pulsars, which are used by pulsar timing arrays (PTAs) to detect gravitational waves from supermassive binary black holes (for more details see \citealp{shannon15, ferdman10,arzou18}). Despite pulsars exhibiting steep spectra which imply higher flux at lower frequencies \citep{seiber}, a large population of pulsars at lower frequencies are marginalized from the PTA analysis, where dispersion and scattering effects are greatest. These reasons further motivate us to undertake a study of propagation effects with pulsars at low frequencies. In this paper we focus on the scattering effects for a sample of pulsars observed at low frequencies ($10-88$ MHz), using the first station of the Long Wavelength Array (LWA1). We examine the scaling relations for scattering with time and frequency. We model for both the frequency dependence of the scattering time as well as the DM since $\alpha$ depends on both.

This paper has been organized in the following manner. In Section 2 we describe our observations and preliminary data reduction; in Section 3, we describe the scattering analysis methods. In Section 4, we describe our results for our sample of seven pulsars and Section 5 contains a detailed discussion and comparison with previous observations.

\section{Observations}

The LWA1 \citep{Taylor12} is a radio telescope array located near the Karl G. Jansky Very Large Array in central New Mexico. It consists of 256 dual-polarized dipole antennas operating in the frequency range $10-88$ MHz. The outputs of the dipoles can be formed into four fully independent dual-polarization beams such that each beam has two independent frequency tunings (chosen from the range $10-88$ MHz)  with a bandwidth of up to 19.6 MHz in each tuning. The ability of the LWA1 to observe multiple frequencies simultaneously provides a powerful tool for studying frequency dependence of pulsar profiles \citep{Ellingson13}. At these low frequencies, the pulses experience scattering and dispersion effects to a much greater extent than at higher frequency. Thus, the LWA1 can be used to make very precise measurements of these effects for studying the ISM properties.

The LWA Pulsar Data Archive\footnote{https://lda10g.alliance.unm.edu/PulsarArchive/} \citep{stovall15} contains reduced data products for over 100 pulsars (Stovall et al., in prep) observed since 2011.  The data products used for this study are produced by coherently de-dispersing and folding the raw LWA data using DSPSR\footnote{http://dspsr.sourceforge.net/index.shtml}. We used archival observations at four frequency bands: 35.1, 49.8, 64.5, and 79.2 MHz. The archival data have already been corrected for DM effects via coherent de-dispersion, and consist of 4096 phase bins for each of 512 spectral channels. The data are saved in the form of 30-second sub-integrations. 

We excise RFI using a median zapping algorithm that removes data points with intensity more than six times compared to the median within a range of frequency channels. These files are then further reduced in two ways, one is to obtain total average profiles with two to four channels (depending on the signal to noise ratio of a pulsar) for scattering studies and the other with eight channels for obtaining the pulse time of arrivals (TOAs) for measuring the DM variation over time. For scattering study, we reduce the number of phase bins to 256 to smooth the average profiles. These tasks are performed using the PSRCHIVE command \texttt{pam} \citep{van12}.  We also remove the baseline from the observed profiles and then normalize them by their peak amplitude.

\begin{table}[h]
\begin{center}
Selected Pulsars Observed with LWA1\\
\vspace{0.1cm}
\begin{tabular}{lllllll}
\hline
Source & DM$_{LWA1}$ & Period & Distance & $PM_{RA}$ & $PM_{Dec}$\\ 
 & (pc cm$^{-3}$) & (s) & (kpc) & (mas/yr) & (mas/yr) & \\
\hline
B0329+54 & 26.7639(1) & 0.71452 & 1.00 & 17.0(0.3) & -9.5(0.4)\\
B0823+26 & 19.4789(2) & 0.5307 & 0.32 & 61.0(3) & -90.0(3) \\
B0919+06 & 27.2986(5) & 0.4306 & 1.10 & 18.8(.9) & 86.4(0.7)\\
B1822$-$09 & 19.3833(9) & 0.7690 & 0.30 & -13(11) & -9(5) \\
B1839+56 & 26.774(1) & 1.6529 & 2.19 & -30(4)& -21(2)\\
B1842+14 & 41.498(1) & 0.3755 & 1.68 & -9(1) & 45(1)\\
B2217+47 & 43.488607(5) & 0.5385 & 2.39 &-12(8) & -30(6)\\
\hline
\hline
\end{tabular}

\caption[The LOF caption]{ {The list of sources studied in this paper. DM and Period values have been obtained from \cite{stovall15}. Values for the Distance, $PM_{RA}$ and $PM_{Dec}$ have been obtained from ATNF{\footnotemark} pulsar catalogue.}}
\end{center}
\end{table}
\footnotetext{http://www.atnf.csiro.au/people/pulsar/psrcat/ .}

In our preliminary study of scattering, we obtain archival data for seven pulsars (Table 1) for all the available epochs since the commission of LWA1. For each pulsar, we split each frequency band into two channels except for two pulsars: PSR B1822$-$09 and PSR B1839+56. For PSR B1822$-$09 we have used four channels to compare our results with previous studies by \cite{kkumar15}. In case of PSR B1839+56, we have reduced the data to four channels to improve our sample size as the S/N at higher frequency bands (64.5 and 79.2 MHz) is poor. The analyzed frequency range was cut for all the pulsars with regard to full LWA capabilities due to S/N issues coming from the shape of pulsar spectra and/or sensitivity. We list center frequencies used for each pulsar in Table 2. 

These seven pulsars were previously noted to have profile shapes below 100 MHz that are consistent with the effects of interstellar scattering \citep{stovall15} and we follow the same data reduction procedure for all of them (see Section \ref{Analysis}).

\section{Analysis} \label{Analysis}

After obtaining the average profiles for each pulsar, we follow the same technique as reported in \citet[hereafter, KK15]{kkumar15} to model scattering on this dataset. This formulation is based on the simple thin screen model \citep{williamson72}. The observed pulse profile can be expressed as a convolution of the frequency dependent intrinsic pulse $P_{i}(t, \nu)$ with the impulse response, characterizing the pulse scatter broadening in the ISM, $s(t)$ the dispersion smear across the narrow spectral channel D(t), and the instrumental impulse response, I(t). This results in $P(t) = P_{i}(t,\nu) \ast s(t) \ast D(t) \ast I(t)$,
where $\ast$ denotes convolution. Following the same analysis as in KK15, we ignore the effect of I(t) as our instrument is stable in time on the timescale of each observation. D(t) can also be ignored as we use coherent dedispersion which corrects for dispersion smearing in the narrow spectral channel.

\subsection{Intrinsic Pulse Model}
The average pulse profile of a pulsar varies intrinsically with frequency in the number of components, their width, amplitude ratio, and separation between them. Hence, it is important to account for frequency dependent effects. Since the intrinsic pulse profile of a pulsar is unknown, we obtain an expected intrinsic profile model (IPM) at our frequency of interest using higher frequency average profiles for each source in Table 1. We assume the effects of scattering at higher frequencies are too small to affect the pulse shape. Average profiles at multiple frequencies enable us to obtain frequency dependent variation in the parameters affecting the pulse shape. For the IPM, we obtain average pulse profiles at frequencies ranging between $\sim 100 - 410$ MHz from the European Pulsar Network (EPN\footnote{http://www.epta.eu.org/epndb/}). However, if one of the profiles in the above frequency range has poor signal to noise ratio (S/N) or two average profiles are close in frequency (143 and 151 MHz), it is difficult to accurately determine the frequency evolution. In such cases, we either consider no frequency evolution (see Table 2) or include our 79.2 MHz data (highest central frequency in the LWA1 band) depending on its S/N. Differences in these average profiles apart from the intrinsic frequency effects of interest can also stem from different telescopes, instrumentation, observation date, and duration.

We model these profiles using a sum of Gaussians as explained in \citet{Kramer94}. The number of Gaussian components increases with each iteration and is limited to a maximum of five. We use various criteria to limit the number of Gaussian components, for example, the iteration stops when the amplitude of the residual maxima is $< 5\%$ of maximum peak intensity or the chi-square value increases with the addition of a new Gaussian component. 

We assume that the number of components does not change within the selected range of frequencies ($79.2 - 410$ MHz). Once we obtain a set of Gaussian parameters for a profile at one of the frequencies (preferably LOFAR high band since it offers high S/N), we use these parameters as an initial condition and apply them to rest of the frequency profiles. We then fit for changes in the main component width based on a power law in frequency (for the list of parameters, see Table 2), as pulsars have broader pulse profiles at lower frequencies \citep{book1}. We also assume that spacing between the components and their relative amplitudes do not vary within our frequency selection. We consider the radius-to-frequency mapping only for the pulse width. Two pulsars (PSR B0329+54 \& PSR B0919+06) in our sample have more than two components which makes determining the evolution of the component separation very complicated. In order to have consistent analysis we refrain from applying radius-to-frequency mapping for the remaining pulsars.


\begin{table}[h]
\begin{center}
IPM Frequency Parameters \\
\vspace{0.1cm}
\begin{tabular}{llcll}
\hline
Pulsar & Epoch Range & Number of & Frequencies & Frequency\\ 
& MJD & Gaussian Components & Used (MHz) & Dependence ($a, b$)\\
\hline
B0329+54 & $57144 - 58254$ & 4 & 40.0, 44.9, 54.7 &  0.00655, -0.0811\\
& & &  59.6, 69.4, 74.3  & \\
&&&&\\
B0823+26 & $57219 - 57899$ & 2 & 44.9, 54.7, 59.6 & ...\\
& & &  69.4, 74.3 & \\
&&&&\\
B0919+06 & $57312 - 58139$& 3 &  44.9, 54.6, 59.6& ...\\
& & & 69.4, 74.3 & \\
&&&&\\
B1822--09 & $ 57232 - 57919 $ & 1 & 47.3, 52.2, 57.2& 0.0188, -0.166\\
& & &  62.0, 67.0, 71.8 & \\
&&&&\\
B1839+56 & $57225 - 58209$  &1 & 27.8, 32.6, 37.6 & ...\\
&&& 42.4, 47.4, 52.2 & \\
&&& 57.2, 62.0& \\
&&&&\\
B1842+14 & $57242 - 58210$ & 2 & 49.7, 59.6, 69.4 & 0.461, -0.917\\
&&& 74.2, 84.1 & \\
&&&&\\
B2217+47 & $57242 - 58085$ & 1 & 44.9, 54.6, 59.6, & 0.127, -0.602\\
& & &  69.4, 74.3, 84.1 & \\
\hline
\hline
\end{tabular}
\caption{This table lists the range of epochs that have been included in this study for each pulsar; the number of Gaussian components; list of frequencies used for each pulsar, and frequency modeling parameters for the main component width to obtain the IPM given by $a \times \nu^{b}$ (discussed in Section 3.1).}
\end{center}
\end{table}

\subsection{Pulse Broadening}

From the high-frequency models described above, we derive the frequency dependence and a new IPM is obtained for all the LWA1 center frequencies separately. The IPM is convolved with an exponential function with a scattering time ($\tau_{sc}$) to obtain a template. This new template is a function of relative flux, phase offset, and scattering timescale. We then use a least square fitting algorithm to fit the template to the observed pulse profiles. This fitting uses the standard deviation of observed pulse profile as the uncertainty and estimates error bars on the fitting parameters. We assume that error on the template is negligible compared to the uncertainty in the data and will not affect the fitting parameters. 
To align the pulse phase of the template and data, we use a low pass filtering technique to smooth the data. To do this, we first calculate the Fourier transform of the average profile. Then we filter out all the frequencies $> 30 \% $ of the maximum frequency (fmax) and rescale the intensities for frequencies between $ 20-30 \% $ of fmax by $(N-n_{i})/N$ where $n_{i}$ is the index of the frequency point and N is the total number of points within the specified frequency range. Intensities at frequency $< 10 \% $ of the fmax are scaled by 1. We include these two frequency ranges to avoid sharp edges. We then take the inverse Fourier transform of the filtered profile and use that to find the peak location. Since this filters out all the high Fourier frequencies, noise interferes less with determining the phase of the peak. 

We plot the fitting examples for PSR B2217+47 at all six frequencies in Figure 1 to demonstrate our procedure. As can be seen, our templates (Figure 1) fit the observed dataset with a reduced chi-square in the range of $1-3$ for all frequencies. Similarly, we have obtained an IPM for all the pulsars in our sample and used them to fit our observations. 

 

\begin{figure}[t!]
\begin{center}
\includegraphics[width=\textwidth,angle=0]{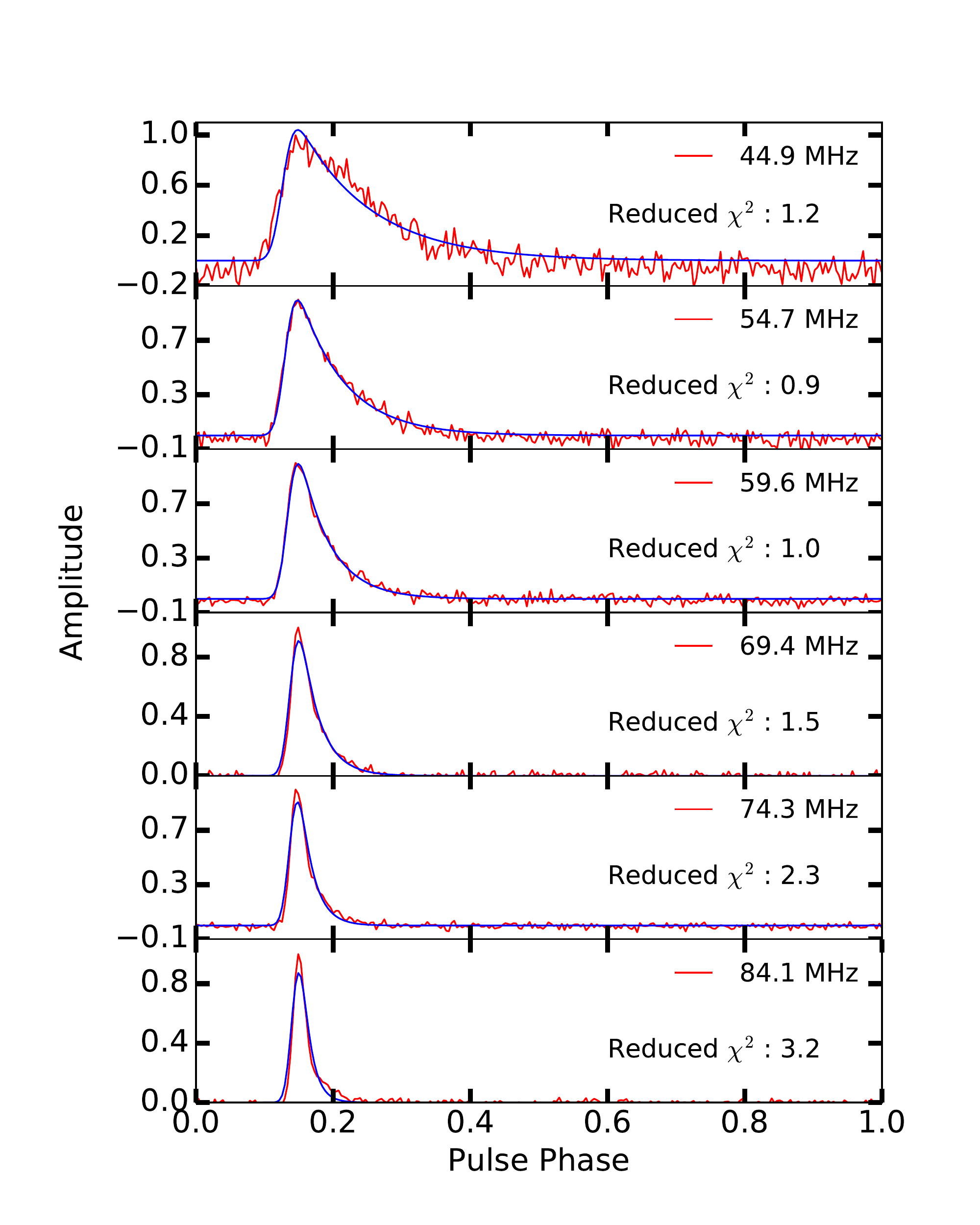}
\end{center}
\vspace*{-0.0in}
\caption{Fitting example for B2217+47 epoch MJD 57372. Top to bottom plots show fitting of intrinsic pulse models convolved with an exponential scattering function to the observed data. For B2217+47, the noise in the data increases as we go towards the lower frequencies.}
\end{figure}

\begin{figure}[t!]
\begin{center}
\includegraphics[width=\textwidth, angle=0]{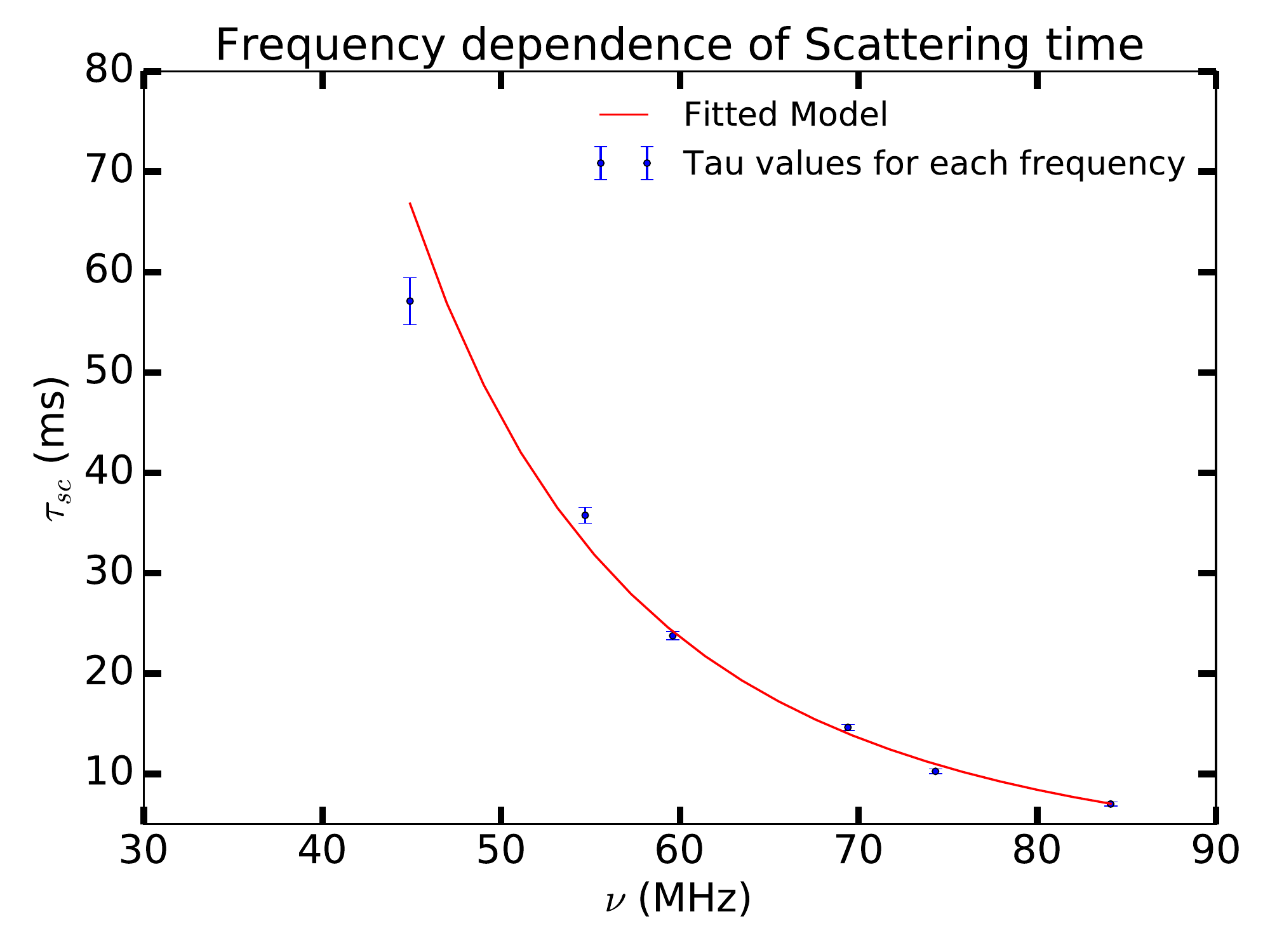}
\end{center}
\caption{Example of $\alpha$ fit for PSR B2217+47 at epoch MJD 57372.}
\end{figure}

\subsection{DM Variation}

Since the scattering time can be related to DM (see Equation 1 and 2) along the LOS, it is important to study if there are any variations in DM over the duration of our observations. For our study, we measure $\delta$DM using the pulsar timing software \texttt{TEMPO}\footnote{http://tempo.sourceforge.net/} DMX parameters. These measure an offset of DM from a fiducial value for multiple epochs each having a specified time span. We used a time span of about three years. 

For the DM analysis, the number of frequency channels in the archive files is reduced to eight for each epoch at two frequencies, 64.5 and 79.2 MHz, using the PSRCHIVE task \texttt{pam}. We only use the higher frequencies since they have a better S/N. The TOAs for these profiles are obtained using \texttt{pat} (a PSRCHIVE algorithm). For the timing model, we first apply ephemeris changes using the task \texttt{pam} to make sure that all of the epochs have the same ephemeris, and then average profiles across all the epochs for each frequency using \texttt{psradd}, followed by smoothing the profile using \texttt{psrsmooth}. We then align the 64.5 MHz and 79.2 MHz pulse models, before obtaining the TOAs. We have converted the TOAs to the solar system barycentre using the DE405 model \citep{Standish98}. The time difference between the observed TOA and the timing model gives timing residuals for each observation. We combine these timing residuals from both the frequencies into one file and fit for the above-listed parameters. 


 \section{Results}

We have studied scattering in seven pulsars using the LWA1 data. Below we discuss our results for each pulsar individually.
\subsection{PSR B0329+54}

\begin{figure}[t!]
\begin{center}
\begin{tabular}{c}
\includegraphics[width=\textwidth,scale=4,angle=0]{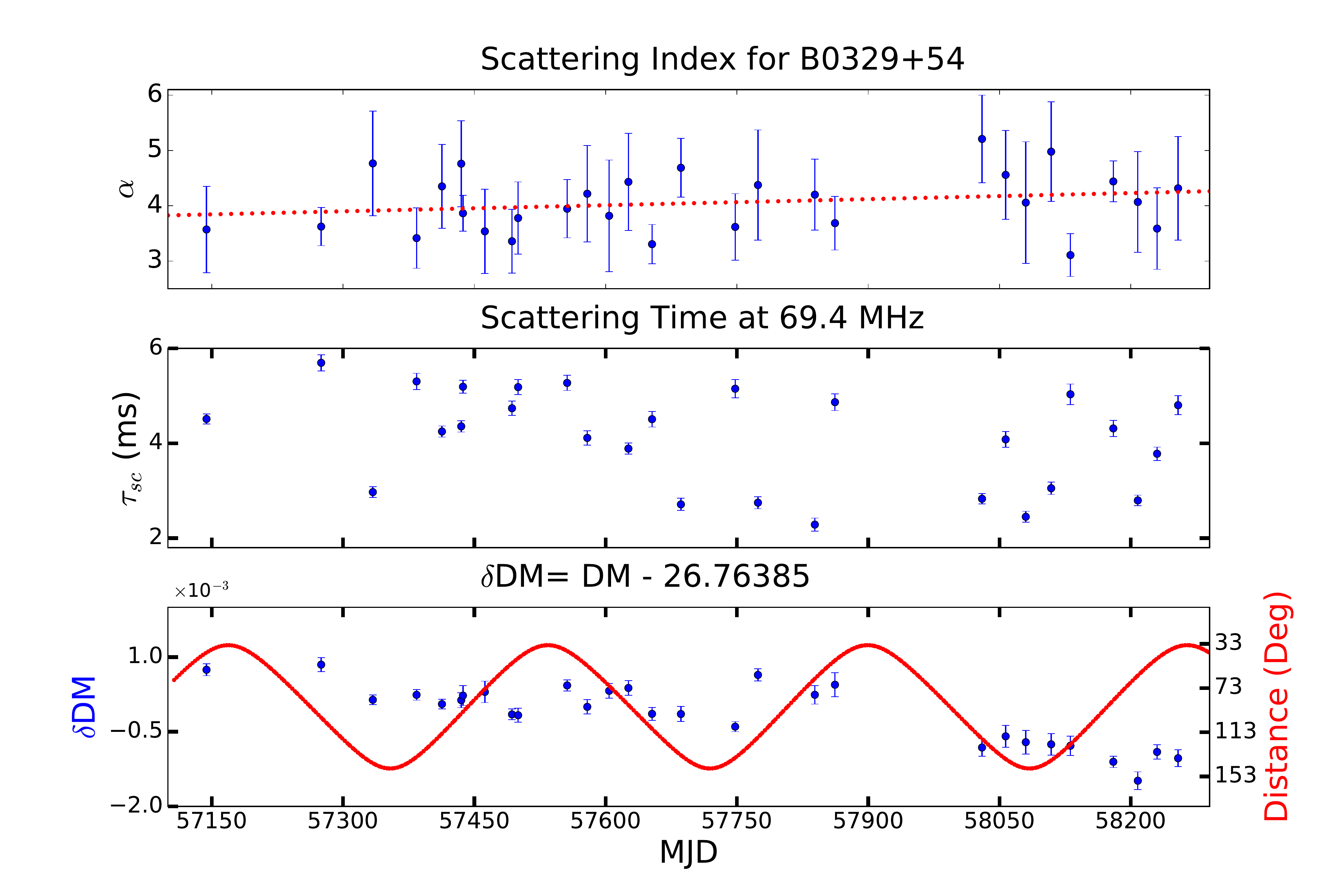}\\
\end{tabular}
\end{center}
\caption{The top and middle panel consist of scattering index values ($\alpha$) and scattering time at 69.4 MHz, respectively, over time for PSR B0329+54. The bottom panel consists of $\delta$DM values (blue) and solar elongation angle (red) over time. All four measurements have been made at the same epochs for a span of about three years. For more details see Section 3.}
\end{figure}

PSR B0329+54 is known to have 9 emission components, one of the highest number of components for any pulsar \citep{ggupta01}. We have used four components where the amplitude of these components is above 5\% of the main component amplitude. The frequency evolution for the main component has been obtained from three frequencies: 79.2, 143, and 408 MHz (obtained from EPN). 

We have plotted $\alpha$, $\tau_{sc}$, $\delta$DM, and the solar elongation angle in Figure 3. The measured median $\alpha$ value is $4.05 \pm 0.14 $. This estimate is in agreement with the prediction of the Gaussian distribution where the $\alpha$ value is expected to be 4.0.  The fitting to the $\alpha$ value over time yields a slope equal to $0.13 \pm 0.11$ year$^{-1}$. This is consistent with there being little or no variation in the scattering index with time.

PSR B0329+54 has been found to have periodic variation in its timing residuals with two different periods likely due to the presence of potential planets \citep{starvo17} in its orbit. This contributes to the timing noise, hence, we have obtained the DM values for this pulsar in a slightly different manner. Instead of fitting for $\delta$DM across all epochs simultaneously, we have fit for each epoch individually. Figure 3(b) shows the variation in $\delta$DM values with an overall change of about 0.0015 pc cm$^{-3}$ over a span of about 3 years. These variations do not correlate with the solar elongation angle (Figure 3c) where the trend is periodic. We do not expect this pulsar to exhibit variation in DM due to the solar wind since the closest distance of approach is $34.2^{\circ}$.This pulsar is also known to exhibit mode switching \citep{chen11}, where the relative amplitude of the component changes and the total pulse width becomes narrower as the profile components change their phases. These mode changes are simultaneous across frequency but non-uniform \citep{bartel81}, hence, will introduce a frequency-dependent variation in timing residuals. Apart from mode change, it is also known to have planets in its orbit which will again affect the timing residuals. Hence, this apparent variation in $\delta$DM is not due to any physical change in ISM, but instead the high timing noise of this pulsar.


\subsection{PSR B0823+26}

PSR B0823+26 pulse profile consists of a main pulse, post-cursor, and an inter-pulse \citep{rankin95}. The amplitude of both inter-pulse and post-cursor is less than 5\% of the main pulse. However, we note there are two additional wing components on both sides of the main pulse that are above 5\% of the main pulse's amplitude at 143 MHz and 151 MHz. It is difficult to obtain frequency dependency from only these two nearby frequencies especially when it involves multiple components. Hence, for the IPM, we have used the profile at 151 MHz and considered no frequency evolution. 

This pulsar is known to exhibit nulling, detected in LOFAR observations \citep{sobey}, which causes several observations without any pulse, hence, have been excluded from this analysis. We have plotted the $\alpha$ and DM values over time in Figure 4 top and bottom plots, respectively. The estimated median $\alpha$ value for this pulsar is $1.55 \pm 0.09$, which is quite small in comparison to the expected theoretical value. We fit a trend to the $\alpha$ values over time and have found the slope value to be $-0.16\pm 0.13$ year$^{-1}$. This is consistent with there being little or no variation in the  scattering index over time. 

From the $\delta$DM  plot (Figure 4b), we see a periodic variation of $\sim 1.7 \times 10^{-3}$ pc cm$^{-3}$ over a period of about half a year. This periodicity in DM can be attributed to the solar wind, due to its close proximity to the Sun (Figure 4c). The closest distance of approach for this pulsar is about $7^{\circ}$. 

\begin{figure}[t!]
\begin{center}
\begin{tabular}{c}
\includegraphics[width=\textwidth,,angle=0]{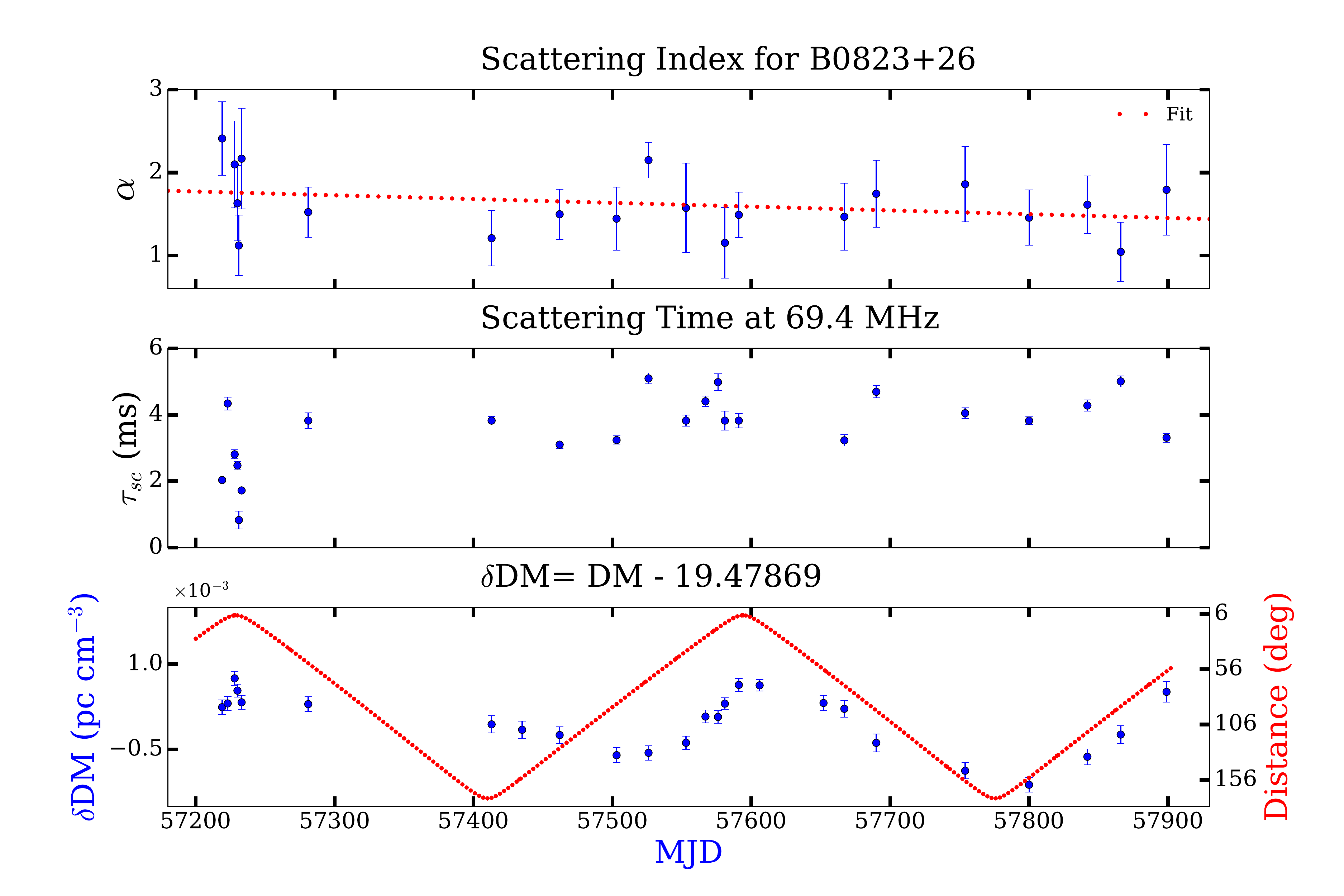}\\
\end{tabular}
\end{center}
\caption{$\alpha$, $\tau_{sc}$ at 69.4 MHz, $\delta$DM, and solar elongation angle for every epoch for PSR B0823+26}
\end{figure}

\subsection{B0919+06}

The number of components in B0919+06 is a matter of some debate since this pulsar is known to exhibit abnormal emission which causes flares a few degrees in phase before the normal emission \citep{rankin06, han16}. The timescale of this emission is about 15 seconds and expected to occur once in 1000 pulses. Since for our analysis we average an hour long archival observation in time, we do not detect it. 
The Gaussian fitting of the 135 MHz profile yields three components and the 408 MHz profile shows only two components, where the post-cursor component is missing. Due to this we only derive the IPM from 135 MHz with no frequency evolution. Since 135 MHz is closer to the frequencies we are working with as compared to 408 MHz which is six times larger. This provides us best possible fits for this pulsar. 

The average $\alpha$ value is $2.88 \pm 0.18$. The slope value is $-0.28 \pm 0.15$ year$^{-1}$. The epochs ranging from MJD 57744 to 57903, either have poor data quality or fewer number of frequencies, hence, have not been included in the fitting plot. For B0919+06, the scattering time ($\sim 5$ ms) is small in comparison to PSR B2217+47 and B0329+49 ($\sim 30$ ms) at the frequencies of LWA1. This could imply that pulsars with similar DM values (PSR B0329+54) have different scattering timescales, as it also depends on the location of a source in the galaxy.

This pulsar appears to have a steep spectrum at the LWA frequencies and, consequently, the S/N deteriorates when we go towards the higher end of the spectrum. Hence, we ignore the last channel (84.1 MHz) for scattering index estimation. For some of the epochs, the average profile fits poorly with the model for the mid-range frequencies and we get a lower scatter timescale for a higher frequency. It seems that this may be attributed to the error in $\tau_{sc}$ and these error bars may be underestimated.

The $\delta$DM plot (Figure 5c) exhibits variation over time with no overall change in $\delta$DM value. The minimum solar elongation angle for this pulsar is $8.36^{\circ}$. Thus, we expect the variation in $\delta$DM due to the solar winds (Figure 5c) to be periodic which does not align with our $\delta$DM observations. This pulsar is known to show variation in the frequency derivative \citep{perera14}. The observed trend in $\delta$DM can, therefore, be explained as a combined effect of both the solar wind and the varying frequency derivative.

\begin{figure}[h!]
\begin{center}
\begin{tabular}{c}
\includegraphics[width=\textwidth,,angle=0]{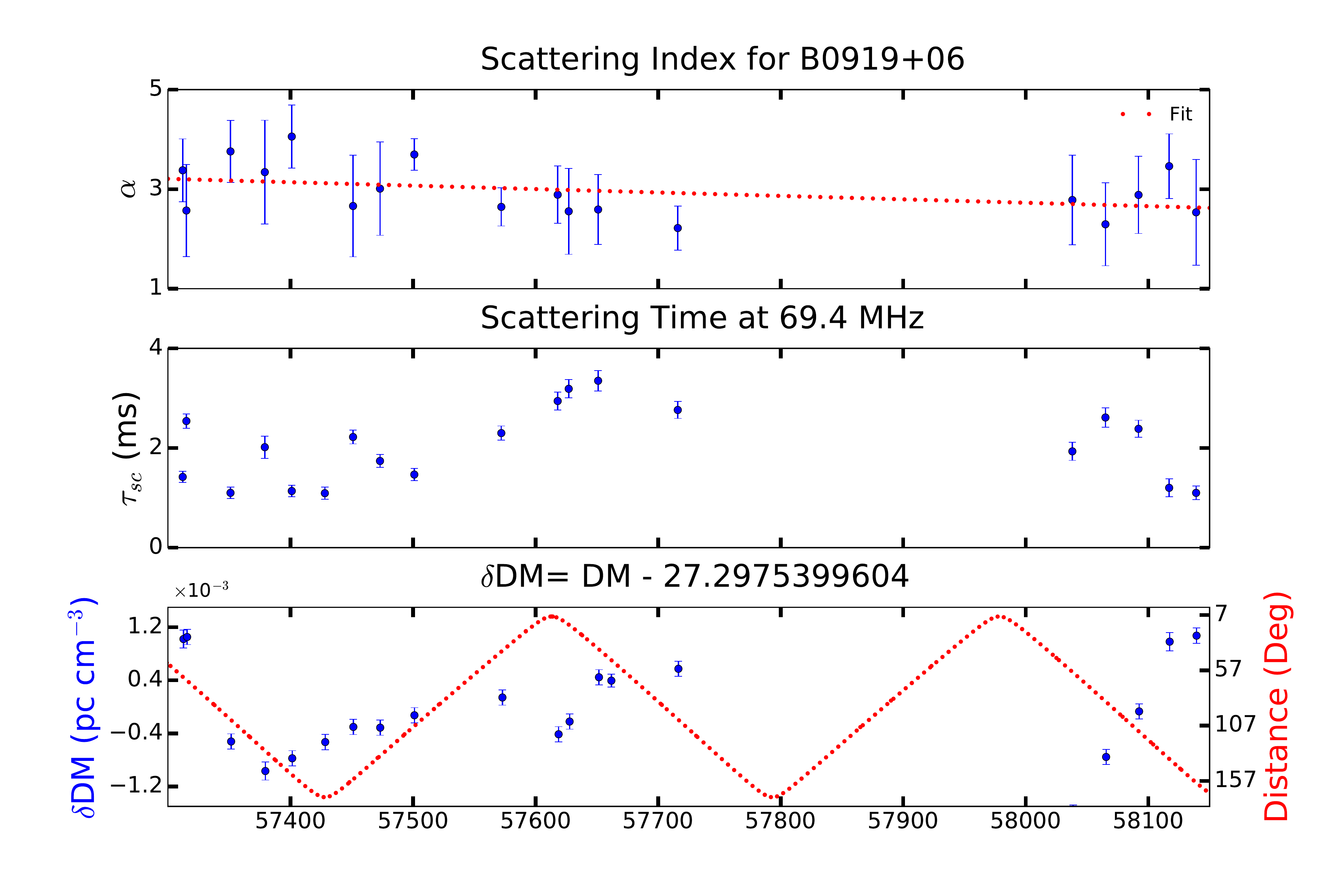}\\
\end{tabular}
\end{center}
\caption{$\alpha$, $\tau_{sc}$ at 69.4 MHz, $\delta$DM, and solar elongation angle for every epoch for B0919+06}
\end{figure}

\subsection{PSR B1822$-$09}

PSR B1822$-$09 is a single component pulsar. We have obtained frequency evolution from these four frequencies: 87, 149, 400, and 408 MHz. For this pulsar, we have reduced the data to four channels instead of 2 to compare our results with \cite{kkumar17} (hereafter, KK17 and see Section 5.1), as they also reduced it to four channels for their analysis. We have plotted $\alpha$, $\delta$DM, and relative solar distance over the observation time in Figure 6. The median $\alpha$ value is $4.18 \pm 0.13$. This value falls in the range of both the Kolmogorov and Gaussian models. The scattering index remains constant over the duration of our observations with a slope equal to $0.11 \pm 0.25$ year$^{-1}$.


PSR B1822$-$09 is a nearby pulsar with a DM of $19.38$ pc cm$^{-3}$ (Table 1). Its minimum solar elongation angle is about 13 degrees. As can be seen in the Figure 6c, there is an overall increment in $\delta$DM value of $4.8 \times 10^{-3}$ pc cm$^{-3}$ over a span of about three years. Additionally, there are smooth rises and dips with a magnitude of $2.4 \times 10^{-3}$ over a span of 167 days, which are due to change in the solar separation angle. The transverse speed of 22 km s$^{-1}$, implying that LOS of this pulsar remains the same. The overall increase in $\delta$DM implies turbulence in the ISM along the LOS. Since the LOS remains the same, the scattering index remains constant

\begin{figure}[t!]
\begin{center}
\begin{tabular}{c}
\includegraphics[width=\textwidth,,angle=0]{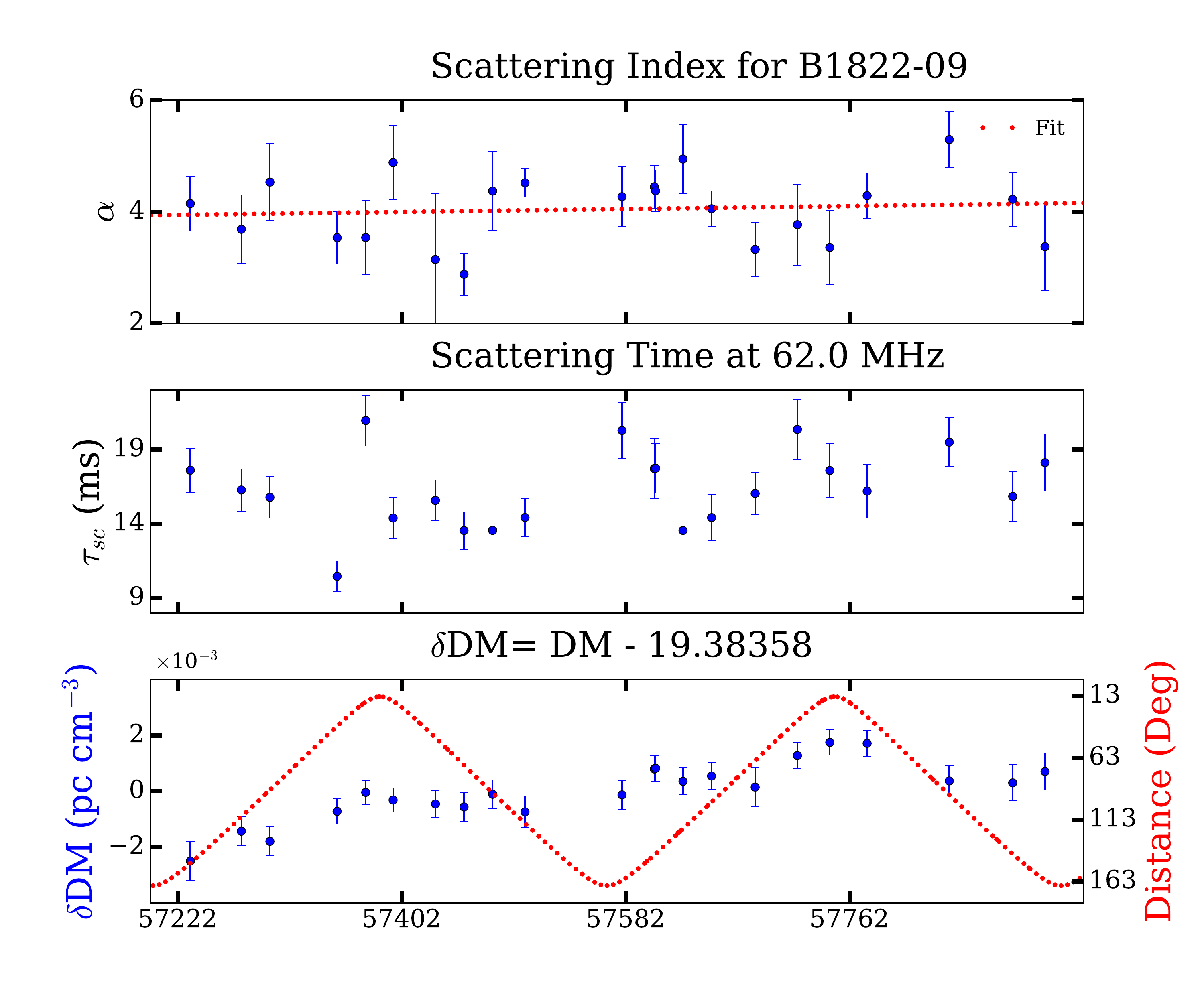}\\
\end{tabular}
\end{center}
\caption{$\alpha$, $\tau_{sc}$ at 62.0 MHz, $\delta DM$, and solar elongation angle for every epoch for PSR B1822$-$09}
\end{figure}

\subsection{B1839+56}
PSR B1839+56 is another single component pulsar. We have derived the IPM using 143 MHz profile. Due to poor S/N at 79.2 MHz, we were unable to obtain frequency evolution for this pulsar. We have plotted the $\alpha$, $\delta$DM  values, and relative solar distance for this pulsar in Figure 7. The median value of $\alpha$ is $2.70\pm 0.16$. The $\alpha$ remain mostly constant with an estimated slope of $0.10 \pm 0.05$ year$^{-1}$.

The $\delta$DM  values for B1839+56 are small compared (of the order of $10^{-4}$ pc cm$^{-3}$) to the other pulsars (of the order of $10^{-3}$pc cm$^{-3}$) discussed in this paper. Also, these error bars are comparable to the variation in $\delta$DM , thus, implying insignificant variation in DM over time. 

\begin{figure}[t!]
\begin{center}
\begin{tabular}{c}
\includegraphics[width=\textwidth,,angle=0]{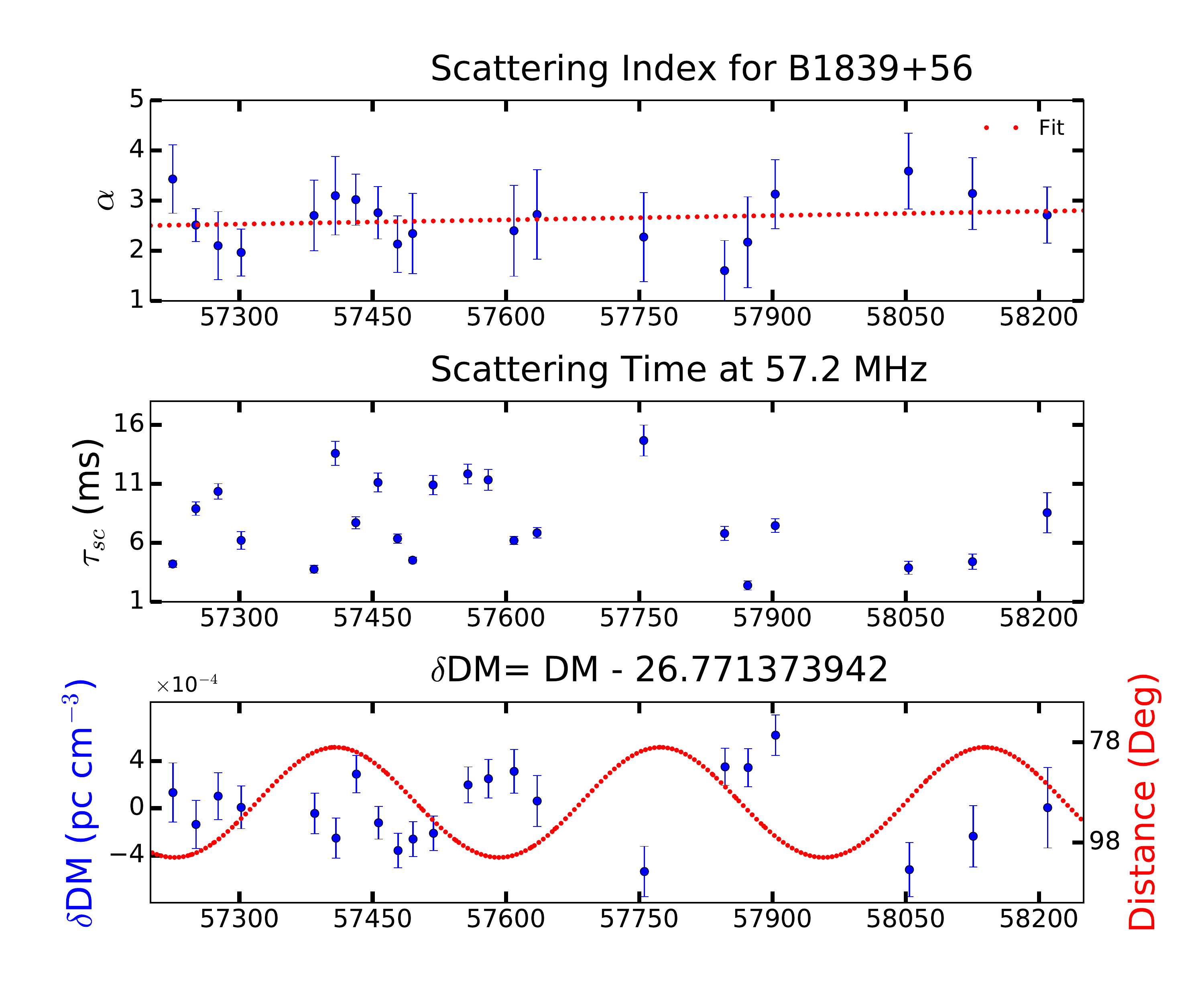}\\
\end{tabular}
\end{center}
\caption{$\alpha$, $\tau_{sc}$ at 57.2 MHz, $\delta$DM, and solar elongation angle for every epoch for B1839+56}
\end{figure}


\subsection{B1842+14}

This is a double component pulsar at LWA frequencies. The IPM has been obtained using the LOFAR data at 143 MHz and the LWA profile at 79.2 MHz. We only use frequency evolution of the main component as discussed in Section 3.1. We have plotted $\alpha$, $\tau_{sc}$ at 69.4 MHz, $\delta$DM, and relative solar distance in Figure 8. The median $\alpha$ value is $3.24 \pm 0.11$. The $\alpha$ remains constant over the duration of our observations with a fitted slope value equal to $0.09 \pm 0.12$ year$^{-1}$. 


From the $\delta$DM  plot (Figure 8c), we see a moderately constant $\delta$DM until epoch MJD 57756 and a linear change of $5.8 \times 10^{-3}$ pc cm$^{-3}$ over 455 days. This does not correlate with the solar elongation angle and the closest angular distance of this pulsar to the Sun is about $37^{\circ}$, which is too far away to affect the DM significantly. The transverse velocity of this pulsar is about 365 km s$^{-1}$ (Table 1), which would affect the LOS, hence, the $\delta$DM. We note that this change in LOS does not affect scattering index.  

\begin{figure}[t!]
\begin{center}
\begin{tabular}{c}
\includegraphics[width=\textwidth,,angle=0]{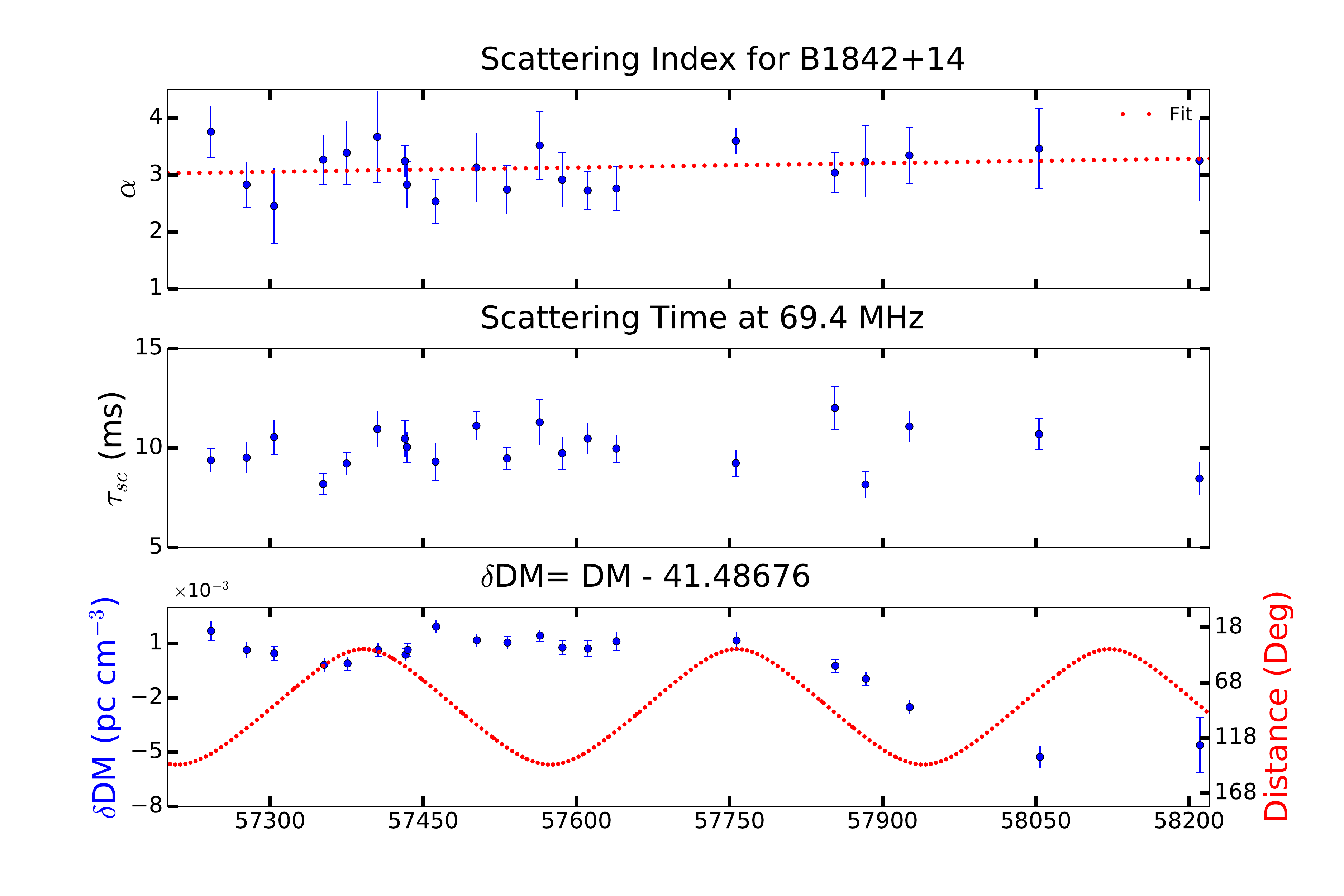}\\
\end{tabular}
\end{center}
\caption{$\alpha$, $\tau_{sc}$ at 69.4 MHz, $\delta$DM, and solar elongation angle for every epoch for B1842+14}
\end{figure}

 \subsection{B2217+47}



PSR B2217+47 is typically known to have a single component below 300 MHz \citep{kuzmin98}, however, recently it was found to have an additional component \citep{pilia16}. This component changes its relative position to the main component over time \citep{michilli}. Since we do not have all the frequency data for the same epoch, we choose to ignore this component and only use the main component for IPM. We obtain the frequency evolution of the pulse component using 79.2, 143, and 151 MHz. This pulsar also has the highest S/N in our sample \citep{stovall15}. 

For PSR B2217+47, we have plotted $\alpha$ values for all epochs to see if there is any variation over time (Figure 9a). The median $\alpha$ is $3.58 \pm 0.10$. The green dotted line represents the linear fit with an estimated slope of $-0.44 \pm 0.10$ year$^{-1}$, which implies a decrement in $\alpha$ at a level of $4.4 \sigma$. 

Figure 9c shows a variation in $\delta$DM  values over time. The overall variation in $\delta$DM  is about 0.005 pc cm$^{-3}$ over a span of 661 days. To understand the variation in DM, we plot the solar elongation of this pulsar during our observation period (Figure 9c), which changes periodically, different from both $\alpha$ and $\delta$DM  trends. The closest angular distance of PSR B2217+47 from the Sun is about $50^{\circ}$ which does not affect the pulsar DM significantly. Since both scattering index and $\delta$DM values vary with time, we test if they are correlated using Spearman rank-order correlation. We estimate the correlation index between the $\alpha$ and $\delta$DM values equal to $-0.56$ (P-value = 0.003), implying a negative correlation between them. We also estimate the correlation index between $\delta$DM  values and scattering timescales at multiple frequencies. For the higher frequency bands, there is a positive correlation with the similar magnitude as $\alpha$, thus confirming our expectation that variation in DM affects the scattering timescale (see Equation 1 and 2). 


\begin{figure}[t!]
\begin{center}
\begin{tabular}{c}
\includegraphics[width=\textwidth,angle=0]{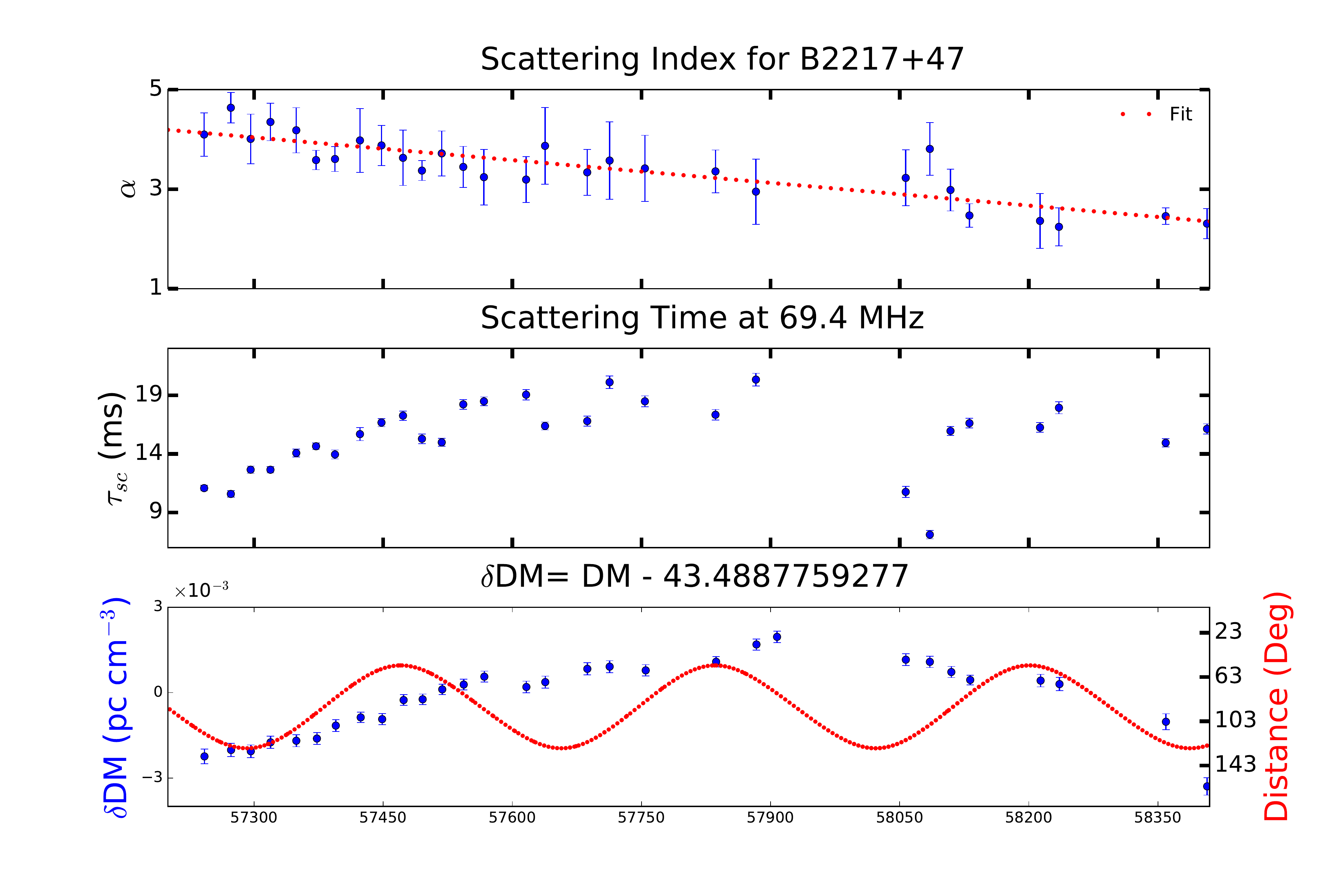}\\
\end{tabular}
\end{center}
\caption{$\alpha$,$\tau_{sc}$ at 69.4 MHz, $\delta$DM, and solar elongation angle for every epoch for B2217+47. We include scattering time to show its correlation with $\delta$ DM (see section 4.7).}
\end{figure}

\section{Discussion \& Conclusion}
We will now summarize our results and discuss what information is gained from time evolution of scattering parameters. We will also discuss how our results align with the thin screen model and the underlying assumptions. 

\subsection{Scattering Spectral Index Distribution}
 
Assuming that the scattering time follows a power law (see Equations 1 and 2), we performed a weighted least-squares fit to estimate the spectral indices for the pulsars in our sample. Our results show that $\alpha$ values except for two pulsars in our sample show deviations from the theoretical power law. Scattering spectral index allows us for an estimate of electron density index ($\beta$; see Equation 4). Table 3 summarizes the scattering spectral index median value and $\beta$ for our sample of pulsars. Since Equation 4 is only applicable for $\alpha > 4$, we have been able to obtain this only for PSR B0329+54 and PSR B1822$-$09 (see \citealp{xu17} for more details).

The $\alpha$ measurements for PSR B0329+54 and PSR B1822$-$09 are consistent with the theoretical value of 4.4 for a Kolmogorov distribution. Our median $\alpha$ value for these pulsars slightly differs from $\alpha$ values of $4.3 \pm 0.1$ and $5.0 \pm 0.5 $, respectively, as reported in KK17. The reason for this discrepancy is likely a result of a different IPM. They have used the 87 MHz profile with no frequency evolution for the IPM whereas we have used a higher frequency (151 MHz) with frequency evolution (refer Section 4.3 and 4.4). We also note that assuming no frequency evolution overestimates the $\tau_{sc}$ and hence, the value for $\alpha$.

PSR B0823+26 has the lowest value of $\alpha$, $1.55 \pm 0.40$ among all the pulsars. We compare this observation with scattering measurement done by \cite{kuzmin07} where they report the $\alpha$ value for this pulsar is $3.68$ (no error bar reported). This is quite different from our observation, however, their value was obtained using only two frequencies, which makes it less reliable. Another way of obtaining scattering spectral index is by using spectral dependence of the decorrelation bandwidth, as done by \citet{daszuta13} for PSR B0823+26. The decorrelation bandwidth is related to the scattering time by $2 \pi \tau_{sc} \delta \nu_{d} = C_{1}$, where $\nu_{d}$ is the decorrelation bandwidth. They have found this value to be $3.94 \pm 0.36$, which also differs significantly from our observation. However, this decorrelation bandwidth measurement is at higher frequencies ($> 300$ MHz) which raises a question of if the scattering index varies with frequency.

Other pulsars with a large $\alpha$ deviation from theoretical expectations are B1839+56 and B0919+06 with estimated median values of $2.70 \pm 0.16$ and $2.88 \pm 0.18$, respectively. We compare our $\alpha$ measurement for B0919+06 with \cite{kuzmin07}, where they have reported a value of $3.05 \pm 0.08$. This agrees with our measurement within the error bars, and compared to PSR B0823+26 is likely to be more reliable since it was obtained using three frequencies.

PSR B2217+47 has a median $\alpha$ value of $3.58 \pm 0.10$. This is lower than the theoretical value for a Gaussian distribution of 4, but closer to that theoretical expectation than $\alpha$ values for four pulsars in our sample. Similar to PSR B0329+54 and B1822$-$09, our $\alpha$ value for B2217+47 agrees with values reported in the literature ($4.2 \pm 0.1$, see KK17) but is slightly higher than our median due to a different IPM. For PSR B1839+56 and PSR B1842+14, this is the first time scattering spectral index has been estimated. In this study, we have considered no frequency evolution for three pulsars (for example PSR B0823+26, see Table 2) in our sample for which we suggest that the reported values should be on the higher end.

\begin{table}[h]
\begin{center}
Scattering Results\\
\vspace{0.1cm}
\begin{tabular}{lcc}
\hline
Pulsar & $\alpha$ & $\beta$\\ 
\hline
B0329+54 & $4.05 \pm 0.14$ & 3.95\\
B0823+26 & $1.55 \pm 0.09$ & ..\\
B0919+06 & $2.83 \pm 0.18$ & ..\\
B1822$-$09 & $4.18 \pm 0.13$ & $3.83$\\
B1839+56 & $2.70 \pm 0.16$ & ..\\
B1842+14 & $3.24 \pm 0.11$ & ..\\
B2217+47 & $3.58 \pm 0.10$ &  ..\\
\hline
\hline
\end{tabular}
\caption{$\alpha$ and $\beta$ values are median values obtained in this study. $\alpha$ and $\beta$ are scattering index and electron density index, respectively ( refer Equation 3 and 4).}
\end{center}
\end{table}

\subsection{Deviation from theoretical models}

PSR B0823+26 is the nearest source in our sample and shows large deviations in the scattering index from the thin screen model. There are three possible explanations for this observation. First, this pulsar shows intrinsic variation such as nulling, sub-pulse drifting, and mode switching, which may affect the average profile and hence, the evaluation of scattering time. This pulsar also has a small scattering time in comparison to PSR B1822$-$09 even though both sources have the same DM. The second explanation for this observation is that since scattering time is small in comparison to the pulse width, it is difficult to obtain the true scattering time and hence, we get a flatter index. The third explanation is based on inner scale effects. Since at lower frequencies the diffraction scale can become smaller than the inner scale, which causes a flatter spectra as compared to the theoretical model. This explains the discrepancy in observed $\alpha$ value with that reported in \citet{daszuta13} since they had observed it at a higher frequency, implying steeper spectra.

As mentioned in the introduction, deviation from the theoretical models has been observed in many scattering observations across all ranges of DM, more often for high DM pulsars \citep{lewan13,lewan15}. KK17 claim that $\alpha$ estimates for DMs $< 50$ pc cm$^{-3}$ are in good agreement with those expected for a Kolmogorov spectrum with an average value of $3.9 \pm 0.5$. However, on an individual basis these sources show a deviation from the expected value.

There are various plausible explanations for these deviations. First of all, the thin screen model assumes an infinite thin screen. This assumption seems valid in the case of medium-range DMs. For large DMs, the probability of having multiple screens increases, hence, the assumption of the thin screen becomes less valid. In the literature, most of the scattering studies \citep{lewan13,lewan15} have been conducted for high DM pulsars (DM$> 300$ pc cm$^{-3}$), where $\alpha < 4$ have been attributed to multiple finite scattering screens. For the low DM sources, the scattering screen can be finite which can lead to $\alpha$ values less than or equal to 4.0 \citep{cordes01}. Spectra with $\alpha < 4$, are expected as a result of finite scattering screen by \citet{rickett09} and anisotropic scattering mechanisms by \cite{stine01}. 

In a recent simulation study, \cite{geyer16} find that for anisotropic scattering, spectral index of scattering time is $< 4$ with the infinite and finite screen. Anisotropy in scattering implies elongated scattering angle in one direction as compared to other direction. However, the effect of anisotropy would be more distinctive in the case of image broadening as compared to temporal broadening. We note that in case of PSR B2217+47 where we see a change in scattering index with time, its velocity along Declination is only 2.5 times the RA which makes it difficult to interpret if the observed trend is due to change in anisotropy. We discuss this source further in more detail in Section 5.3. In order to conclusively determine the effect of anisotropy, in addition to measuring the temporal broadening, we either need to image the source simultaneously or study its dynamic spectrum. From this study, we suggest that low DM pulsars do not always follow Kolmogorov distribution or the Gaussian distribution. It is likely due to the deviation of the scattering model from the thin screen model even for the nearby pulsars.

Apart from DM, another important factor that will likely affect the scattering is the location of a pulsar in the Galaxy. Among the previous studies, \citet*{lewan15} report no correlation between the scattering spectral index and the distance or the position of the source in our Galaxy. Similarly, KK17 do not suggest any trend in their data with respect to the location in our Galaxy and emphasize that to draw conclusions from such a plot, we need DM-independent distance measurements.






\subsection{Time Evolution of Scattering parameters}
The main focus of this paper is to study the evolution of $\alpha$ with time and to understand it in more detail we also obtain variation in DM in parallel for the same epochs. In our sample of seven pulsars, only B2217+47 shows a significantly varying $\alpha$, whereas we see a variation in DM for all pulsars except PSR B1839+56. This variation in DM can be periodic, linear or a combination of both depending on the underlying effects. A linear trend in DM can be explained either due to the change in LOS due to the proper motion of a pulsar or a change in distance to the screen along the LOS \citep{petroff13, lam15}. Periodic variation in DM can be attributed to either ionosphere or the solar wind. We estimated the contribution of the ionosphere to DM which is of the order of $< 10^{-4}$ pc cm$^{-3}$, about 10 times smaller than the DM variations we measured, hence would not affect our observations. We used an IONEX Global Ionosphere Model\footnote{ftp://cddis.gsfc.nasa.gov/gnss/products/ionex/} to model the slant total electron content in the ionosphere and converted that to a DM since they are the same measurement with different units. From our observations we note that solar wind will affect the DM for the slow pulsars with a minimum solar elongation of about $15^{\circ}$.

For PSR B0823+26, the $\delta$DM plot shows periodic variation, which is mainly due to the change in solar elongation angle. Similarly, for B1822$-$09, we see smooth periodic rises and dips due to change in angular distance to the Sun, as well as an overall change of $0.001$ pc cm$^{-3}$ per year in $\delta$DM. This implies a slight variation in the column density along the LOS.

We note that we have derived these DM values from timing solutions which also depend on the intrinsic properties of a pulsar including the ISM. In such cases, the derived DM values may not represent the actual DM of the ISM. This is true for PSR B0329+54 and B0919+06 for which intrinsic variations in pulsar cause large timing residuals. The $\delta$DM plot for PSR B0329+54 shows variation in $\delta$DM, however, the actual variation is in the timing residuals due to the possibility of planets in its orbit \citep{starvo17}. The $\delta$DM variation for B0919+06 also includes variation in pulsar frequency derivative. Since the trend in $\delta$DM for both pulsars are due to intrinsic reasons, this requires further investigation and is beyond the scope of this paper. This, however, is beyond the scope of this analysis. 

We expect a change in DM due to the change in LOS will also affect the scattering spectral index. This is the case of PSR B2217+47 which shows a decrement in $\alpha$ values with time. The $\alpha$ and $\delta$DM are anti-correlated and the scattering time at higher frequencies is positively correlated with $\delta$DM (Figure 9). This pulsar is known to have an additional component, a trailing component in the main pulse \citep{michilli}. However, we do not see this additional component at our frequencies. We see a similar change in DM as reported in \cite{michilli} of $0.004$pc cm$^{-3}$ over the same duration. This implies that this pulsar is encountering a gas cloud with a higher density and its structure is changing, which is affecting the pulse broadening.

We note that variations in DM may not affect the scattering spectral index if the motion is parallel to the LOS since the ISM structure would remain the same along the LOS \citep{lam15}. In the case of PSR B1842+14, there is a change in DM but no variation in the scattering index. This implies that this change is likely due to a change in electron density along the LOS and the structure of ISM remains the same.


\section{Summary}
We present a study of scattering spectral index and DM variation for seven pulsars over the timescale of $\sim 3$ years using the LWA1. This is the first time a systematic evolution of $\alpha$ has been reported according to our survey of the literature. Most of the pulsars in our sample exhibit constant $\alpha$ throughout the observations with the slope value consistent with zero. The exception to this is PSR B2217+47 where we measure a decrement in $\alpha$ over our observation period of three years which anti-correlates with a change in DM. For PSR B0823+26, we obtain the smallest $\alpha$ value of $1.55 \pm 0.40$, indicating inner scale effects. 

The median scattering spectral index for five of the seven studied pulsars is below 4, implying deviation from both Gaussian and Kolmogorov inhomogeneities for DM $< 50$ pc cm$^{-3}$. $\alpha$ measurements at lower frequencies and their deviations from theoretical models have led to an improved understanding of correlations between ISM structure and pulsar scattering, but the detailed structure of the ISM and the physical interpretation still remain unclear. Anisotropy is another likely explanation for these deviations in scattering spectral index. However, to effectively understand the anisotropy in the ISM, we need to study the dynamic spectra parallel to the temporal broadening for a larger sample of pulsars. More observations of other pulsars at these low frequencies will be helpful in understanding the distribution of scattering spectral index with DM. Similarly, DM-independent distance measurements will be helpful in obtaining the scattering index distribution across the Galaxy.
 
 \section{Acknowledgements}
We would like to thank Joe Malins to help with the ionosphere DM estimations. We also thank the anonymous referee for valuable suggestions that has improved this paper. Construction of the LWA has been supported by the Office of Naval Research under Contract N00014-07-C-0147 and by the AFOSR. Support for operations and continuing development of the LWA1 is provided by the Air Force Research Laboratory and the National Science Foundation under grants AST-1835400 and AGS1708855.





\begin{thebibliography}{natbib}
\bibliographystyle{apj}
\bibitem[Abott et al.(2016)]{abott16}B. P. Abbott et al., PRL 116, 061102, 2016
\bibitem[Arzoumanian et al.(2018)]{arzou18} Arzoumanian, Z. and Baker, P.T. and Brazier, A. et al., ApJ, 2018, 859, 47
\bibitem[Bartel et al.(1981)]{bartel81}Bartel N., Morris D., Sieber W., Hankins T., 1982, The Astrophysical Journal, 258, 776
\bibitem[Chen et al.(2011)]{chen11}J. L. Chen, H. G. Wang,et al., ApJ, 2011, 741, 48
\bibitem[Cordes \& Lazio(2001)]{cordes01}Cordes J., Lazio T., 2001, ApJ, 549, 997
\bibitem[Cordes \& Lazio(2002)]{ne2001}Cordes J., Lazio T., 2002, arXiv:astro-ph/0207156
\bibitem[Daszuta et al.(2013)]{daszuta13}M. Daszuta, W.Lewandowski J. Kijak, arXiv:1310.8076v1
\bibitem[Ellingson et al.(2013)]{Ellingson13}S. W. Ellingson, G. B. Taylor, J. Craig, et al., 2013, IEEE Transactions on Antennas and Propagation, 61, 2540 
\bibitem[Ferdman et al.(2010)]{ferdman10}Ferdman, R.D., van Haasteren, R., Bassa, C.G. et al., 2010, arXiv:1003.3405
\bibitem[Gangadhara and Gupta(2001)]{ggupta01}R. T. Gangadhara 1 and Y. Gupta, 2001, ApJ, 555, 31
\bibitem[Geyer \& Karastergiou(2016)]{geyer16}Geyer M., Karastergiou A., 2016, MNRAS, 462, 2587
\bibitem[Geyer et al.(2017)]{geyer17}M. Geyer, A. Karastergiou, V.I. Kondratiev et al., arXiv:1706.04205v1
\bibitem[Han et al.(2016)]{han16}Han, J., Han, J.L., Peng, L.-X et al. 2016, \mnras, 456, 3413
\bibitem[Krishnakumar et al.(2017)]{kkumar17}M. A. Krishnakumar1,2,3, Bhal Chandra Joshi2, and P. K. Manoharan, 2017, ApJ, 846, 104
\bibitem[Krishankumar et al.(2015)]{kkumar15}M. A. Krishnakumar, D. Mitra, A. Naidu, et al., 2015, Astrophysical Journal, 804:23 (9pp)
\bibitem[Kramer (1994)]{Kramer94} M. Kramer, 1994, \aaps, 107, 527
\bibitem[Kuzmin et al.(1998)]{kuzmin98}Kuzmin, A. D., Izvekova, V. A., Shitov, Y. P., et al. 1998, A\&AS, 127, 355
\bibitem[Kuzmin \& Losovsky(2007)]{kuzmin07}Kuzmin A. D., Losovsky B. Y., 2007, Astronomical and Astrophysical Transactions, 26, 597
\bibitem[Lambert \& Rickett(1999)]{lambert99}H. C. Lambert AND B. J. Rickett, 1999, ApJ, 517, 299
\bibitem[Lyne \& Graham Smith(2006)]{book1}Pulsar Astronomy by Andrew Lyne and Francis Graham-Smith (Cambridge: Cambridge Univ. Press)
\bibitem[Lam et al.(2016)]{lam15} M. T. Lam, J. M. Cordes, S. Chatterjee, et al., arXiv:1512.02203v2
\bibitem[Lewandowski et al.(2013)]{lewan13}Lewandowski, W., Dembska, M., Kijak, J., \& Kowalinska, M. 2013, MNRAS, 434, 69
\bibitem[Lewandowski et al.(2015)]{lewan15}Lewandowski, W., Kowalinska, M., \& Kijak, J. 2015, arXiv:1502.06330v2
\bibitem[Lohmer et al.(2001)]{lohmer01}Lohmer, O., Kramer, M., Mitra, D., Lorimer, D. R., \& Lyne, A. G. 2001, ApJ, 562, 157
\bibitem[Lohmer et al.(2004)]{lohmer04}Lohmer, O., Mitra, D., Gupta, Y., Kramer, M., \& Ahuja, A. 2004, A\&A, 425, 569
\bibitem[Michilli et al.(2018)]{michilli} Michilli, D. , Hessels, W., Donner J. Y., astro-ph, arXiv:1802.03473v1
\bibitem[Rankin and Rathnasree(1995)]{rankin95}Rankin and Rathnasree, 1995, Journal of Astrophysics Astronomy, 16
\bibitem[Rankin et al.(2006)]{rankin06} Rankin J. M., Rodriguez C., Wright G. A. E., 2006, MNRAS, 370, 673
\bibitem[Rickett et al.(2009)]{rickett09}Rickett, B. J., Johnston, S., Tomlinson, T., \& Reynolds, J. 2009, MNRAS,395, 1391
\bibitem[Romani et al.(1986)]{romani86}Romani, R. W., Narayan, R., \& Blandford, R. 1986, MNRAS, 220, 19
\bibitem[Pilia et al.(2016)]{pilia16}Pilia M., Hessels J. W. T., et al., 2016, A\&A, 586, A92
\bibitem[Perera et al.(2014)]{perera14}B. B. P. Perera, B. W. Stappers, P. Weltevrede, et al. 2014, \mnras, 446, 1380
\bibitem[Petroff et al.(2013)]{petroff13}E. Petroff, M. J. Keith, S. Johnston et al., 2013, MNRAS, 435, 1610
\bibitem[Standish(1998)]{Standish98} Standish,E.M.,1998a. JPL Planetary and Lunar Ephemerides, DE405/LE405 A,JPL IOM 312.F-98-048
\bibitem[Stovall et al.(2015)]{stovall15}K. Stovall, P. S. Ray, J. Blythe et al., 2015, arXiv:1410.7422v2 [astro-ph.IM]
\bibitem[Scheuer(1968)]{scheuer68}Scheuer, P. A. G. 1968, Natur, 218, 920
\bibitem[Shannon et al.(2015)]{shannon15} Shannon, R. M., Ravi, V., Lentati, L. T., et al. 2015, Sci, 349, 1522
\bibitem[Stinebring et al.(2001)]{stine01}Stinebring D. R., McLaughlin M. A., et al., 2001, ApJ Letters, 549, L97
\bibitem[Sieber(1973)]{seiber}Seiber W., 1973, A\&A, 28, 237
\bibitem[Sobey et al.(2015)]{sobey} C. Sobey, N. J. Young, J. W. T. Hessels, et al., arXiv: 1505.03064v1
\bibitem[Starovoit \& Rodin(2017)]{starvo17}E. D. Starovoit and A. E. Rodin, 2017, Astronomy Reports, 61, 948
\bibitem[Taylor et al.(2012)]{Taylor12} Taylor, G. B., Ellingson, S. W., Kassim, N. E., et al. 2012, Journal of Astronomical Instrumentation, 1, 50004
\bibitem[van Straten et al.(2012)]{van12}van Straten W., Demorest P., Oslowski S., 2012, Astronomical Research and Technology, 9, 237
\bibitem[Williamson (1972)]{williamson72} Ian P. Williamson 1972, \mnras, 157, 55
\bibitem[Xu et al.(2017)]{xu17}Siyao Xu and Bing Zhang, 2017, ApJ, 835
\bibitem[Yu et al.(2017)]{ne2017} J. M. Yao, R. N. Manchester and N. Wang, 2017, ApJ, 835
\end{thebibliography}
\end{document}